\documentclass[11pt]{article}

\usepackage[final]{acl}
\usepackage[table]{xcolor}
\usepackage{times}
\usepackage{latexsym}

\newcommand{\gcheck}{\textcolor{green!60!black}{\ding{51}}}
\usepackage{amssymb}
\usepackage{booktabs}
\usepackage{pifont}

\newcommand{\baseRow}{\rowcolor{blue!6}}
\newcommand{\newRow}{\rowcolor{orange!10}}
\usepackage{multirow} 
\usepackage{makecell}
\usepackage[most]{tcolorbox}
\usepackage[T1]{fontenc}
\usepackage{algorithm}
\usepackage{algpseudocode}
\usepackage{subcaption}
\tcbuselibrary{breakable}
\newcommand{\gcross}{$\times$}
\newcommand{\hlblue}[1]{%
  \begingroup\setlength{\fboxsep}{1pt}%
  \raisebox{0pt}[0pt][0pt]{\colorbox{blue!6}{\strut #1}}%
  \endgroup
}
\newcommand{\hlorange}[1]{%
  \begingroup\setlength{\fboxsep}{1pt}%
  \raisebox{0pt}[0pt][0pt]{\colorbox{orange!10}{\strut #1}}%
  \endgroup
}

\usepackage[utf8]{inputenc}

\usepackage{microtype}
\usepackage{hyperref}
\usepackage{inconsolata}

\usepackage{graphicx}

\usepackage{enumitem}
\setlist[itemize]{leftmargin=*, labelsep=0.5em}

\title{Design First, Code Later: Aesthetically Pleasing Template-Free Slides Generation}

\author{
 \textbf{Zhiyao Cui}\textsuperscript{1,2,3}\footnotemark[2],
 \textbf{Chenxu Wang}\textsuperscript{2,4}\footnotemark[2],
 \textbf{Shuyue Hu}\textsuperscript{2},
 \textbf{Yiqun Zhang}\textsuperscript{2}
\\
 \textbf{Wenqi Shao}\textsuperscript{2,3},
 \textbf{Qiaosheng Zhang}\textsuperscript{2,3}\footnotemark[1],
 \textbf{Zhen Wang}\textsuperscript{1}\footnotemark[1]
\\
\\
 \textsuperscript{1}School of Cybersecurity, Northwestern Polytechnical University,
\\
 \textsuperscript{2}Shanghai Artificial Intelligence Laboratory,
\\
 \textsuperscript{3}Shanghai Innovation Institution,
\\
 \textsuperscript{4}Fudan University
\\
 \small{
   \href{mailto:email@domain}{\{cuizhiyao,wangchenxu,hushuyue,zhangyiqun,shaowenqi,zhangqiaosheng\}@pjlab.org.cn,w-zhen@nwpu.edu.cn}
 }
}

\begin{document}

\maketitle
\footnotetext[2]{Co-first, equal contributions.}
\footnotetext[1]{Corresponding authors.}

\begin{abstract}

Producing presentation slides automatically entails coordinating narrative structure with page-level graphic design under strict spatial constraints. For such structured multimodal tasks, a well-organized design process is essential to ensure the final quality of slides. Existing approaches rely on fixed templates or directly emit executable code, thereby both limiting the creative layout-design capabilities of LLMs and bypassing the essential slide‑page design step. To address these limitations, this paper:
(1) proposes a hierarchical slides generation workflow \textbf{DeepSlides} that systematically organizes slide design tasks without any predefined template or style, decoupling slide-page design from implementation;
(2) introduces \textbf{SlideDesign}, a dataset tailored specifically for slides generation tasks;
(3) presents a multi-agent reinforcement learning training paradigm and trains a couple of models \textbf{SlideQwens} for slide design and implementation.
Experimental results demonstrate that our proposed framework outperforms baseline methods on evaluated metrics and achieves superior performance in human preference evaluations.
The dataset and code are available at: \url{https://github.com/sxswz213/DeepSlides}

\end{abstract}

\section{Introduction}

\begin{figure}[ht]
  \centering
  \includegraphics[width=0.95\linewidth]{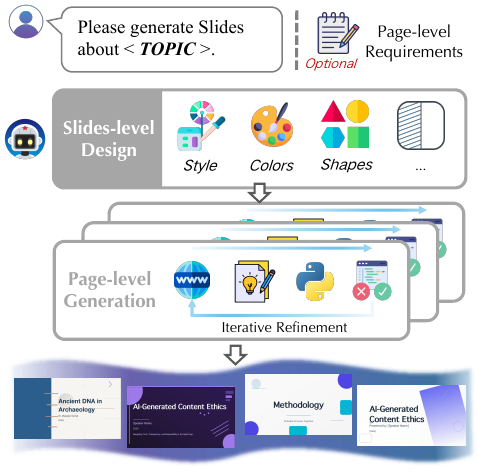}
  \caption{Overview of the DeepSlides workflow. Given a topic and optional user-specified page-level requirements, the system first determines the high-level slide style. Subsequently, each slide is generated through iterative refinement.
}
  \label{fig: overview}
  \vspace{-4mm}
\end{figure}

\begin{figure*}[ht]
  \centering
  \includegraphics[width=0.98\linewidth]{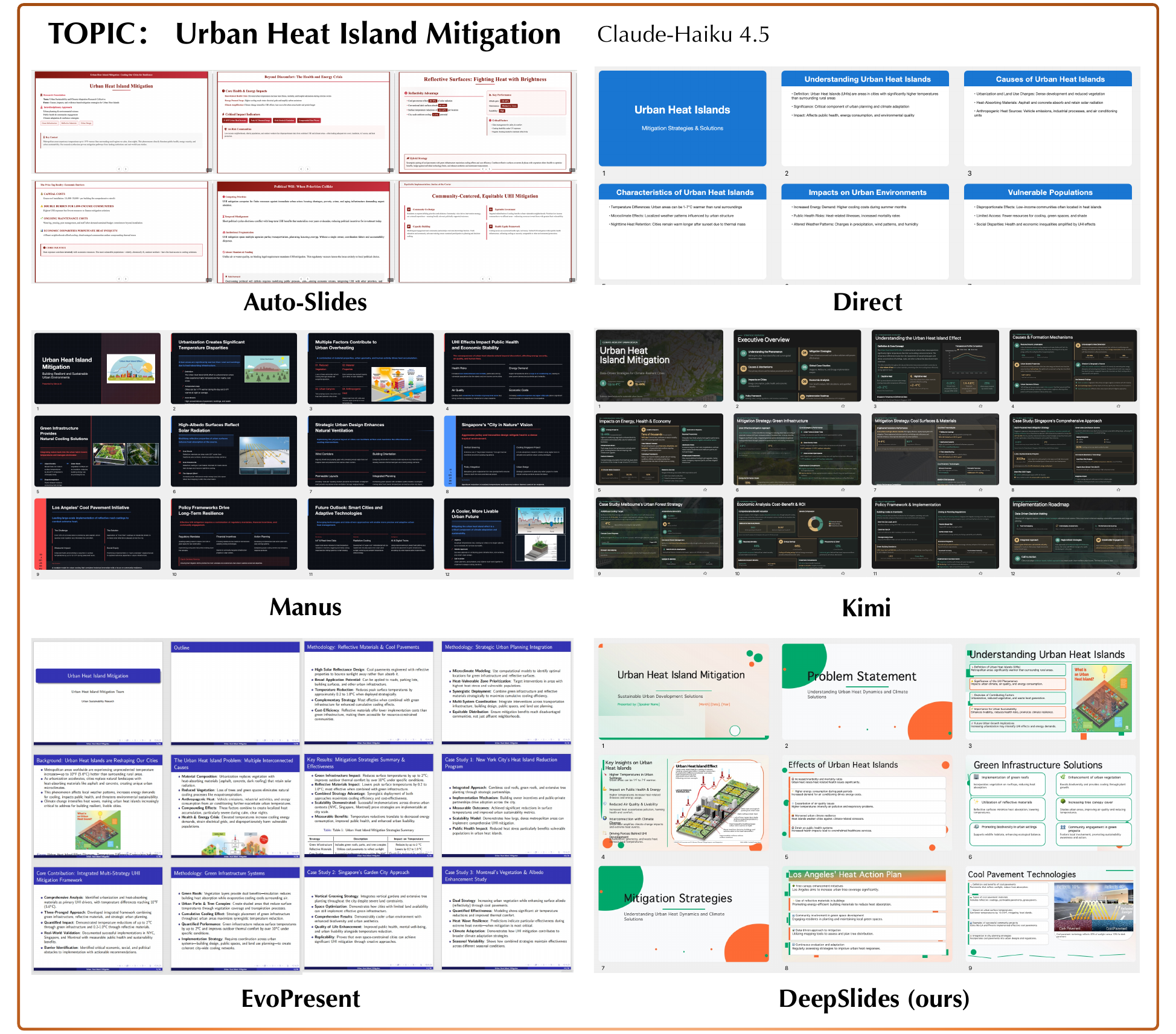}
  \caption{Comparative visualization of slide decks generated for the topic "Urban Heat Island Mitigation", contrasting slides produced by direct model generation, Auto-Slides, EvoPresent, DeepSlides (ours), and two commercial solutions, Manus and Kimi. Additional cases can be found in Appendix~\ref{sec:examples}.}
  \label{fig:case}
    \vspace{-4mm}
\end{figure*}

With the rapid advancement of large language models (LLMs), an increasing number of LLM-based agents are capable of automating various tasks, significantly reducing human workloads~\cite{hong2023metagpt,wang2025perpilot}. In particular, multimodal LLMs possess robust creative abilities and multimodal perception~\cite{meng2025mmeureka}, prompting researchers to explore their application in automated slide generation tasks. As shown in Table~\ref{tab:system-compare}, Google's NoteBookLM~\cite{google_notebooklm} utilizes text-to-image models to produce visually impressive slides; some other platforms, such as Kimi~\cite{moonshot_kimi} and Manus~\cite{manus_ai}, employ HTML-based approaches or predefined templates to ensure slide stability. Further research has investigated methods for transforming image-rich documents or research papers into slide presentations~\cite{doc2ppt_aaai2022, pptagent_2025,liu2025presentingisanart}, enabling incremental edits to existing slides under human supervision~\cite{jung2025talk,guo2024pptc}, or generating individual slides directly through executable code~\cite{ge2025autopresent,tang2025slidecoder}.

Despite the impressive reasoning and content generation abilities of recent LLMs, the visual quality of automatically generated presentations remains far from satisfactory. Existing systems often produce slide decks that are \textit{visually dull}, \textit{compositionally weak}, and \textit{lack expressive power}, failing to capture audience attention in real-world professional settings.  A high-quality presentation should function as a sequence of visually engaging scenes, in which layout choices, visual elements, and graphic compositions contribute to the clarity and effectiveness of the presentation.
As illustrated in Figure~\ref{fig:case}, Auto-Slides~\cite{yang2025autoslides} rely heavily on rigid, template-driven layouts; Manus~\cite{manus_ai} prioritizes decorative aesthetics while neglecting semantic grounding; Kimi~\cite{moonshot_kimi} generates abundant content but suffers from limited visual and structural coherence; and EvoPresent~\cite{liu2025presentingisanart}, although academically rigorous, adopts an overly dense and heavy presentation style. Collectively, these approaches either overemphasize surface-level decoration or focus primarily on content generation, yet still fail to deliver presentations that are visually coherent, creatively composed, and engaging for real-world use.

\begin{table}[tbp]
\vspace{-6mm}
\centering
\small
\setlength{\tabcolsep}{1mm}
\begin{tabular}{lccc}
\hline
& \makecell{Output\\Format}
& \makecell{Complete\\Presentation}
& \makecell{Template\\Free}
\\
\hline
\baseRow EvoPresent   & HTML & \gcheck & \gcross \\
\baseRow AutoSlides   & LaTeX/PDF & \gcheck & \gcross \\
\baseRow SlideGen     & PPTX & \gcheck & \gcross \\
\baseRow SlideCoder   & PPTX & \gcross & \gcross \\
\baseRow DOC2PPT      & PPTX/PDF & \gcheck & \gcross \\
\baseRow PPTAgent     & PPTX/HTML & \gcheck & \gcross \\
\baseRow PASS         & PPTX & \gcheck & \gcross \\
\baseRow AutoPresent  & PPTX & \gcross & \gcheck \\
\midrule
\newRow  NotebookLM   & Image & \gcheck & \gcheck \\
\newRow  Kimi         & PPTX/Image & \gcheck & \gcheck \\
\newRow  Manus        & PPTX/PDF & \gcheck & \gcheck \\
\midrule
\baseRow \textbf{DeepSlides} & PPTX/Image & \gcheck & \gcheck \\
\bottomrule
\end{tabular}
\caption{Comparison of slide-generation workflows. \hlblue{light blue} rows indicate open-source methods, while \hlorange{light orange} rows denote closed-source systems.}
\label{tab:system-compare}
\vspace{-6mm}
\end{table}

To address these limitations, we introduce \textbf{DeepSlides}, a hierarchical, design-first framework that conceptualizes slide generation as the synthesis of a sequence of visually coherent scenes, where layout structure, visual elements, and graphic composition collaboratively convey semantic content and direct audience attention. At the core of DeepSlides is a principled decoupling between design reasoning and code realization. The designer module operates within a high-level semantic design space, responsible for orchestrating visual hierarchy, spatial rhythm, typographic flow, and chromatic harmony across scenes without being constrained by low-level PPTX implementation details. In contrast, the coder module is dedicated solely to compiling these finalized design specifications into stable and executable slide code. This modular separation mitigates the distortion of aesthetic intent by implementation artifacts, limits error propagation across stages, and supports the independent optimization of visual expressiveness and execution robustness, thereby enabling DeepSlides to produce coherent, professionally composed slide decks that go beyond prevailing content-centric, template-constrained, or decoration-driven generation paradigms.

Our empirical results demonstrate that DeepSlides consistently outperforms existing baseline methods across key evaluation metrics, with clear advantages in \textbf{visual layout, hierarchy construction, and color scheme design}. We conducted a human preference and style evaluation with 20 volunteers, where all methods were presented anonymously in random order to avoid bias. Compared with open-source baselines~\cite{yang2025autoslides,liu2025presentingisanart}, DeepSlides achieves a win rate of \textbf{76.5\%}, exceeding the second-best method (9.5\%) by a margin of \textbf{68.0\%}. Even when compared with commercial systems~\cite{manus_ai,moonshot_kimi}, DeepSlides attains a win rate of \textbf{52.0\%}, outperforming the strongest commercial baseline (19.5\%) by \textbf{32.5\%}. These results indicate that DeepSlides is consistently perceived as more visually attractive and creative by human evaluators. In the human style evaluation, DeepSlides obtains strong scores across all four dimensions, including clarity, creativity, visualization and aesthetics, further confirming its overall presentation quality.

To further verify the quality of data produced by our workflow, we construct SlideDesign, a multidisciplinary multimodal dataset under the Field of Science (FOS) taxonomy, which provides paired textual and visual resources for training and evaluation. Based on this dataset, we train a suite of lightweight models on the automatically generated data. As a result, even with a 0.6B-parameter backbone, the resulting SlideQwens models achieve good performance in slide design and code generation, thereby demonstrating the effectiveness of our workflow.


\begin{figure*}[ht]
  \centering
  \includegraphics[width=0.99\linewidth]{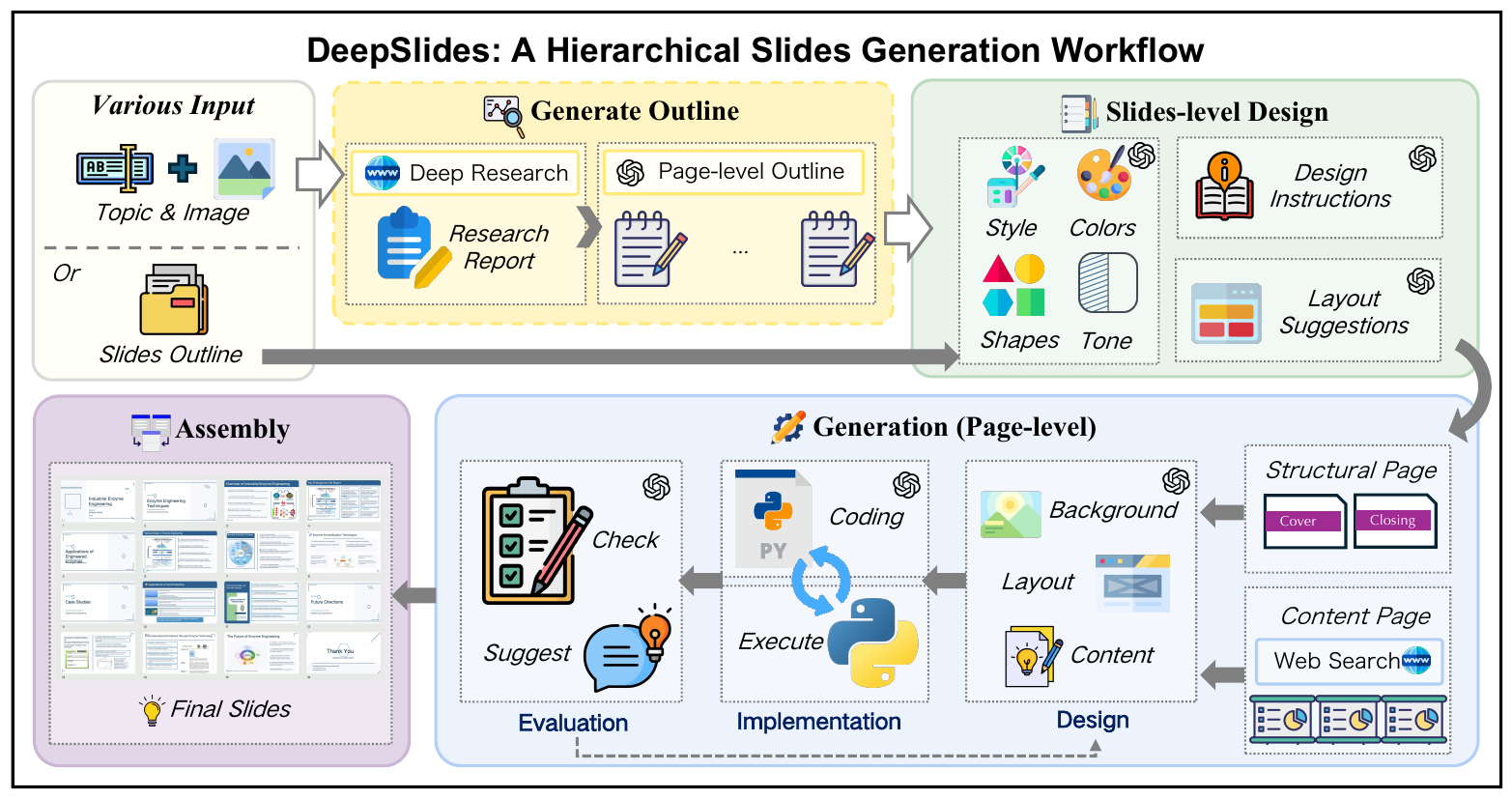}
  \caption{Illustration of the DeepSlides hierarchical slides generation workflow. The framework begins by generating a detailed outline through multimodal deep research or directly from user-provided outlines. Subsequently, slide-level style guidelines are defined, including colors, shapes, and backgrounds. At the page-level, the system performs content expansion via web search and executes detailed design across background, layout, and content layers. Python code is generated to implement each slide, followed by iterative refinement based on evaluation and feedback. Finally, completed slides are assembled into a cohesive presentation.}
  \label{fig: framework}
    \vspace{-4mm}
\end{figure*}

\section{Related Work}

\subsection{Slides-level generation workflows}
While earlier studies had already explored automatic slide generation~\cite{sun2021d2s}, the rapid progress of LLMs and VLMs has dramatically advanced the field, enabling increasingly effective agentic workflows for slide generation~\cite{mondal2024presentations}. PPTAgent~\cite{pptagent_2025} decomposes the task into reference analysis and edit actions, while PASS generalizes automated deck creation from Word documents~\cite{aggarwal2025pass}. SlideGen coordinates collaborative multimodal agents with iterative refinement and layout planning~\cite{liang2025slidegen}, and Auto-Slides~\cite{yang2025autoslides} employs a multi-agent paper-to-deck pipeline with interactive refinement and content verification. In parallel, EvoPresent~\cite{liu2025presentingisanart} focuses on iterative self-improvement via an aesthetic-aware feedback model for multi-round refinement.

Despite strong progress, existing workflows often either (i) rely on template-driven generation, restricting the design space, or (ii) output LaTeX/HTML-style slides that ease rendering but sacrifice native PowerPoint editability and fine-grained interactive, drag-and-drop visual editing.

\subsection{Page-level generation, editing and training}
A growing body of work targets \emph{single-slide} generation and editing via executable representations to preserve editability. AutoPresent~\cite{autopresent_cvpr2025} introduces SlidesBench and shows that programmatic slide synthesis improves controllability and quality over end-to-end image generation. SlideCoder~\cite{tang2025slidecoder} addresses reference-image-to-slide generation with a Slide Complexity Metric and a layout-aware, RAG-enhanced code synthesis framework.

Complementary efforts study \emph{in-place editing} within real presentation software. Talk-to-Your-Slides enables instruction-driven PowerPoint editing with a planning--execution architecture and releases TSBench~\cite{jung2025talk}. PPTC and PPTC-R benchmark multi-turn PowerPoint task completion and robustness to instruction and software shifts~\cite{guo2024pptc,zhang2024pptcr}, while PPTArena and PPTBench further evaluate realistic, layout-sensitive edits and generations~\cite{ofengenden2025pptarena,huang2025pptbench}.

From a training perspective, AutoPresent and SlideCoder primarily target code generation. However, design is typically optimized only implicitly through program correctness. 

\section{DeepSlides}
\label{sec:workflow}
In this section, we first introduce the overall DeepSlides workflow. Given a topic and optional user requirements (e.g., preferred style, color schemes), the system first produces high-level guidance for slides style and layout. It then iteratively refines the content, page design, implementation, and feedback-based revision for each individual slide, and finally aggregates the generated pages into a complete PPTX file. The prompts used in the workflow are provided in Appendix~\ref{sec:prompts}.

\subsection{User Requirement Simulation}

In real-world presentation authoring, the process is typically initiated by a domain expert with substantial background knowledge. The expert initiates the process by constructing a high-level outline, which involves determining the necessary sections, allocating an appropriate number of slides to each section, and specifying the core content of each slide. Only after this process is completed does the expert start authoring individual slides.

To endow the agent with expert‑level background knowledge, we introduce a \emph{DeepResearch} module that guides structured knowledge acquisition and integration, enabling the agent to synthesize web search results and retrieved images into multimodal research reports. Based on this comprehensive report, the agent subsequently outlines the slide deck through three steps: (1) determining the appropriate sections; (2) allocating slide budgets to each section; and (3) generating titles and concise bullet points for individual slides. Collectively, these steps yield a structured, multi-level slide outline for the entire presentation.

\subsection{Slides-Level Design}

At the Slides level, we focus on two aspects: (i) \emph{inter-page coherence} and (ii) \emph{inter-page diversity}. Inter-page coherence requires all pages in a deck to share a consistent style, ensuring they are clearly perceived as part of the same presentation. Inter-page diversity requires that individual pages adopt layouts that are tailored to their specific content, rather than reusing a single template throughout, which would cause visual fatigue.

\paragraph{Slides features.} First, the agent determines the overall style of the slides according to the topic and user requirements, including the global tone, the main color palette, and the font colors for titles and body text. The agent then refines this design by describing the overall visual style in more detail, such as what shapes or patterns should be used as decorative elements, together with their colors, sizes, and approximate positions.

\paragraph{Style Instructions.} We categorize pages into two types: \emph{content pages} and \emph{functional pages}. Functional pages, including the cover, end page, and section divider pages, primarily organize the logical structure of the presentation and do not contain substantive content. In contrast, content pages carry the core information and detailed material. Based on the previously generated style description, the agent formulates specific design guidelines tailored for both functional and content pages. It further refines the visual design by providing detailed descriptions of decorative elements, such as recommended shapes or patterns, and specifying their colors, sizes, and approximate placements.

\paragraph{Layout Instructions.} Finally, conditioned on the global outline, the agent produces a concise, slide-specific layout specification (e.g., “horizontal blocks”). We perform this step at the slides-level to explicitly promote cross-slide structural diversity while preserving overall stylistic coherence.

\subsection{Page-Level Generation}


For content pages, we additionally introduce a content expansion stage, whereas this step is unnecessary for functional pages. Specifically, given a slide’s bullet points, the agent first generates targeted search queries and conducts web retrieval to gather supplementary textual and visual materials. The retrieved information is then distilled into candidate text snippets and image descriptions to enrich the slide content. Subsequently, we perform per-slide design, implementation, evaluation and feedback.

\paragraph{Design.}
We decompose the design of a slide into three layers: \emph{background layer}, \emph{layout layer}, and \emph{content layer}. The agent designs the page by reasoning over these three layers:

\begin{itemize}[nolistsep, noitemsep, leftmargin=*]
    \item \textbf{Background layer.} This layer includes background images, decorative elements, textures, and other elements that are largely independent of the page’s specific content and layout.
    \item \textbf{Layout layer.} This layer specifies the page layout: the arrangement of structural blocks (e.g., title block, bullet block, image block), their functional roles, spatial positions, and any background panels, borders, or separators associated with these blocks.
    \item \textbf{Content layer.} This layer specifies the concrete textual and visual content to be placed into the structural blocks, including the exact text snippets and images.
\end{itemize}

\paragraph{Implementation.}
Given the page-level design, the agent translates the design specification into Python code that programmatically constructs the slide. The resulting slide is then rendered and converted into an image, which will be used for quality assessment. We adopt Python to generate fully editable PPTX files, which better aligns with typical human workflows, rather than HTML- or LaTeX-based outputs (as used in EvoPresent~\cite{liu2025presentingisanart} and Auto-Slides~\cite{yang2025autoslides}) that are comparatively difficult to edit.

\paragraph{Evaluation and Feedback.}
The agent evaluates each slide along three dimensions: \emph{completeness}, \emph{compliance}, and \emph{aesthetics}.

\begin{itemize}[nolistsep, noitemsep, leftmargin=*]
    \item \textbf{Completeness} measures whether all input requirements are reflected in the design and whether the textual and visual content matches the outline and retrieved material.
    \item \textbf{Compliance} measures whether the designed text and image blocks overlap improperly, whether the occupied area of each block is appropriate, and whether any content spills outside the page.
    \item \textbf{Aesthetics} measures the visual appeal of the slide, including the harmony of color and typography, the clarity of visual hierarchy, and the overall balance and alignment of elements.
\end{itemize}



Based on the resulting score and auxiliary feedback signals, the agent decides whether to accept the current design or to iteratively refine the page, feeding targeted revision suggestions provided by the checker into the next design cycle.

\begin{figure*}[htbp]
  \centering
  \includegraphics[width=1.0\linewidth]{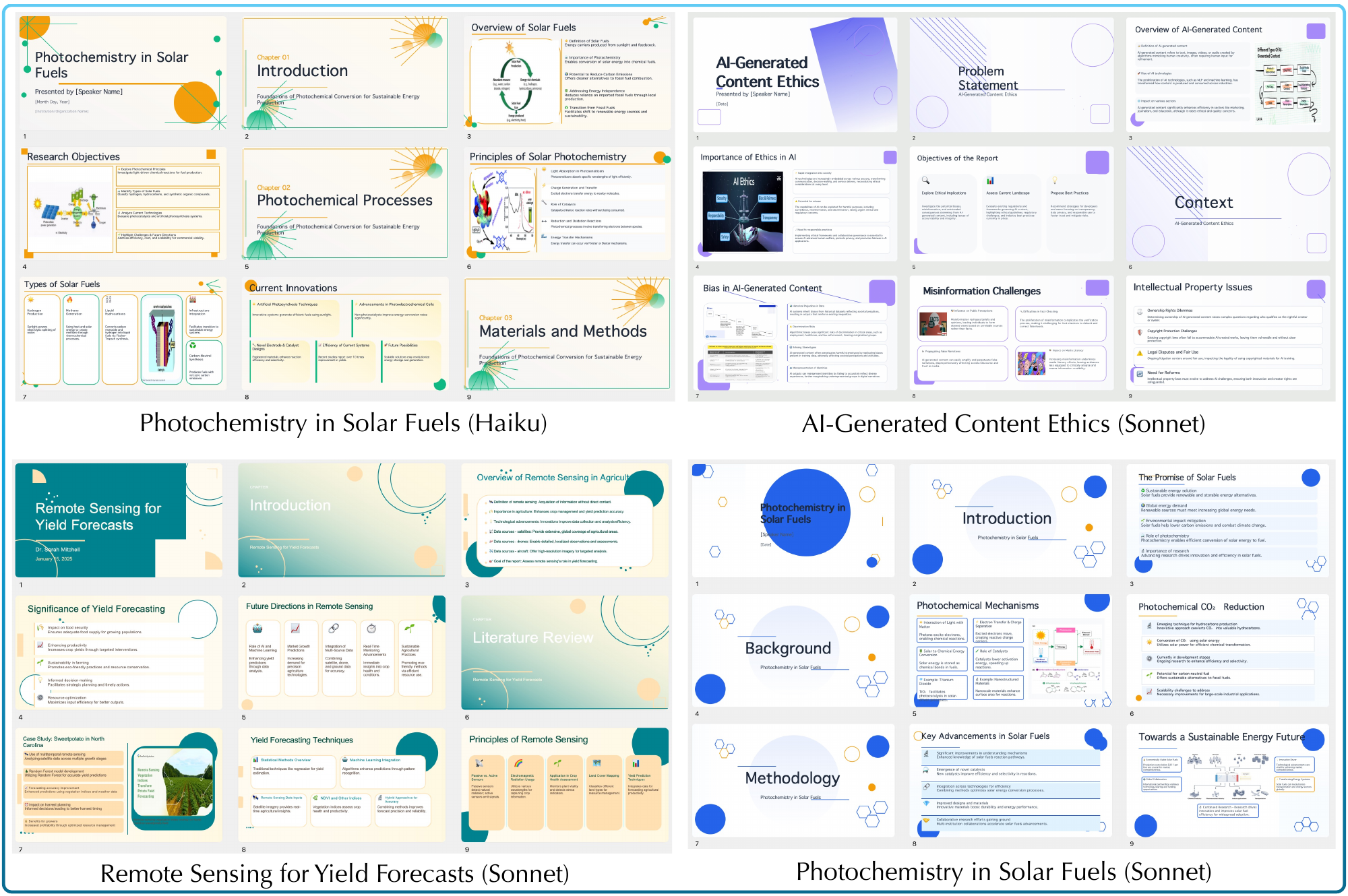}
  \caption{Examples of data in SlideDesign.}
  \label{fig: cases}
    \vspace{-4mm}
\end{figure*}

\begin{figure*}[ht]
  \centering
  \includegraphics[width=1.0\linewidth]{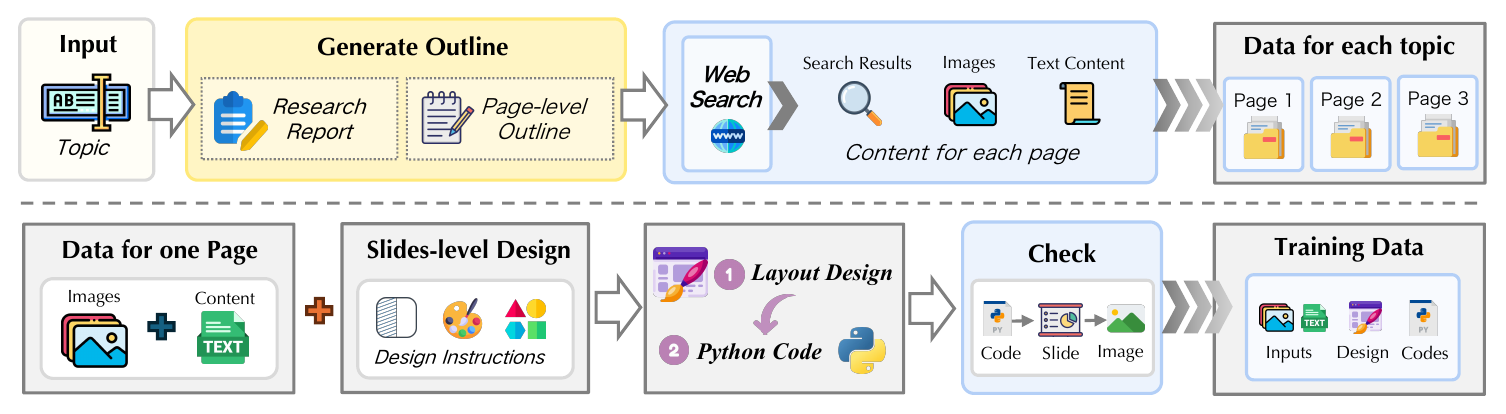}
  \vspace{-4mm}
  \caption{Data generation pipeline }
  \label{fig: pipeline}
  \vspace{-3mm}
\end{figure*}

\begin{figure}[ht]
  \centering
  \includegraphics[width=1.0\linewidth]{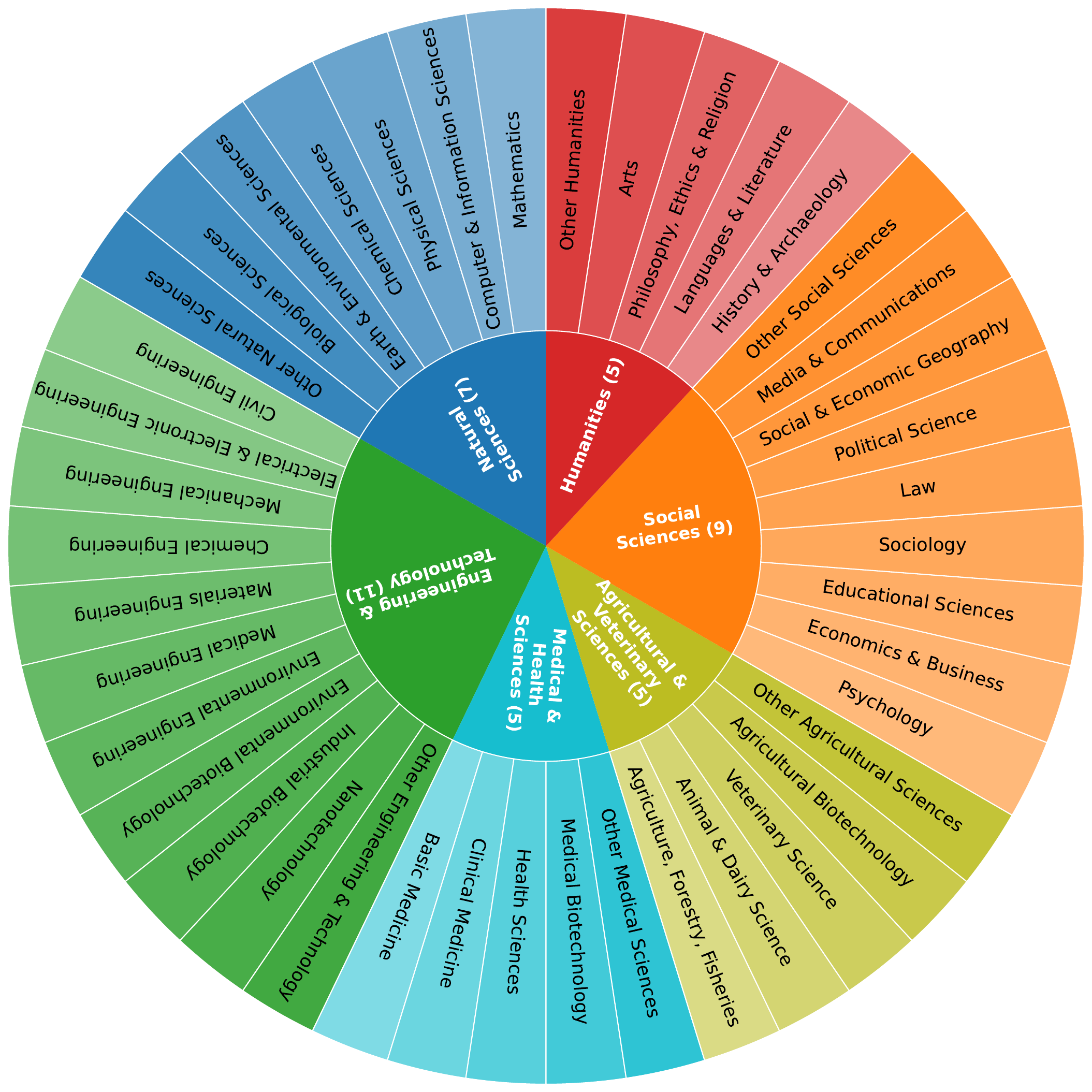}
  \caption{Subject taxonomy and dataset composition.
Inner ring: 6 primary areas with counts; outer ring: 42 secondary disciplines. }
  \label{fig: pie}
\end{figure}

\section{Multi-Agent Training of DeepSlides}
\label{sec:training}


We employed a multi-agent reinforcement learning (MARL) approach to train a set of \textbf{QwenSlides} models, consisting of two specialized models: Designer and Coder. We also construct a dataset, \textbf{SlideDesign}, for slides generation tasks, as shown in Figure~\ref{fig: cases}. 

\subsection{Data Collection}

As shown in Figure~\ref{fig: pipeline}, each topic first goes through outline generation to produce structured slide plans (titles and bullet points). We then run outline-guided web searches to collect supporting text and images, and augment each slide with randomly sampled design specifications (e.g., style preferences and color constraints). Based on these inputs, the SFT pipeline generates paired layout designs and executable code, and retains only samples whose code runs successfully, forming the final training set. We construct the \textbf{SlideDesign} dataset from 420 topics derived from the Field of Science (FOS) classification as shown in Figure~\ref{fig: pie}. More details are reported in Appendix~\ref{sec:dataset}.

\subsection{Training process}

We first perform SFT as a cold start for both the designer and coder, as prior evidence suggests that SFT-based initialization can rapidly adapt models to new tasks and substantially reduce the computational burden of subsequent RL optimization. Building on these SFT checkpoints, we adopt a multi-agent reinforcement learning framework inspired by ReMA~\cite{wan2025rema}, where two agents, $Designer_{\mathrm{sft}}$ and $Coder_{\mathrm{sft}}$, are optimized via interactive rollouts that jointly perform slide layout design and executable code generation. Feedback is provided by a VLM-based reward model that evaluates the outputs along multiple quality dimensions. More training details are reported in Appendix~\ref{sec:training_details}.

\begin{figure}[ht]
  \centering
  \includegraphics[width=0.99\linewidth]{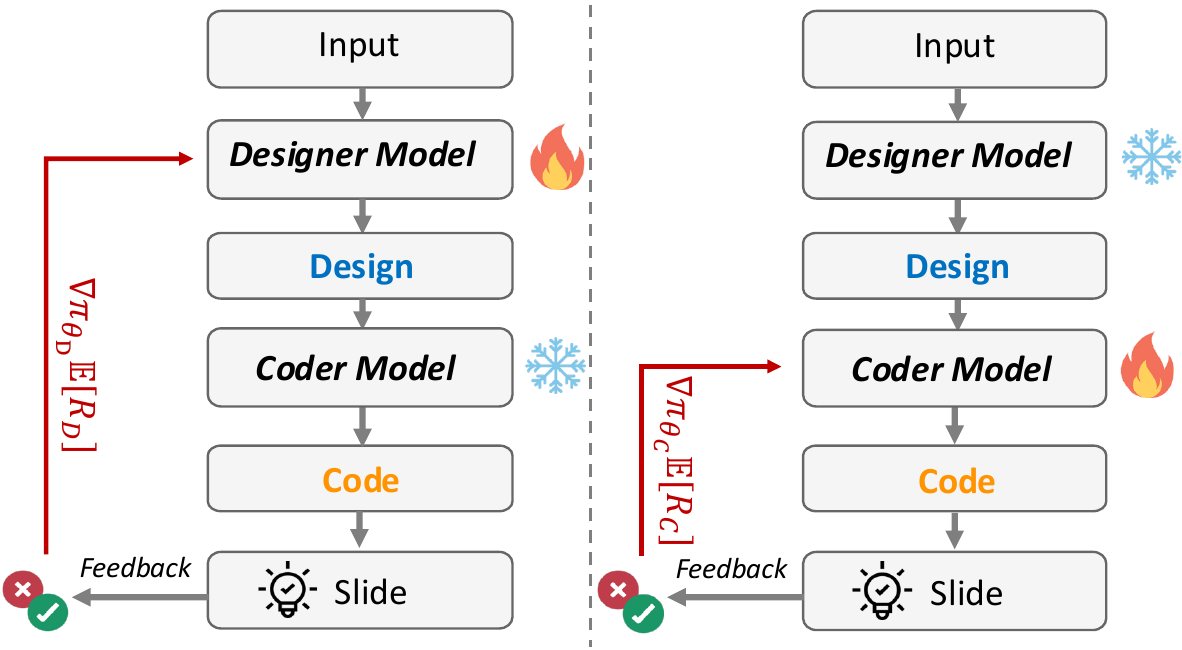}
  \caption{RL train}
  \label{fig: rl}
\end{figure}

Specifically, we first initialize two models ($Designer_{sft}$ and $Coder_{sft}$) with SFT. As illustrated in Figure~\ref{fig: rl}, at each timestep $t$, the designer observes an input prompt $s_t$ and outputs a slide layout design action $a_t^D$ following its policy $\pi_{\theta_D}(a_t^D \mid s_t)$. Subsequently, the coder receives this design as part of its state $s_t'$ and produces the corresponding Python code action $a_t^C$ according to its policy $\pi_{\theta_C}(a_t^C \mid s_t')$. The environment then transitions to the next state, and both agents receive feedback. 

We define the coder reward $R_C$ as the product of (i) executability and (ii) implementation completeness. 
Let $\mathbb{I}_{\mathrm{exec}}\in\{0,1\}$ indicate whether the generated Python code executes successfully, and let $r_{\mathrm{imp}}\in[0,1]$ denote the implementation-completeness score predicted by the reward model (measuring fidelity to the given layout specification). 
The coder reward is
\begin{equation}
    R_C \;=\; \mathbb{I}_{\mathrm{exec}} \times r_{\mathrm{imp}} .
\end{equation}

The designer reward $R_D$ combines completeness, compliance, and aesthetics. 
Let $r_{\mathrm{cmp}}\in[0,1]$ and $r_{\mathrm{cpl}}\in[0,1]$ be the completeness and compliance scores produced by the reward model (cf.\ Section~\ref{sec:workflow}). 
To assess aesthetics, we render the generated slide into an image and obtain a raw aesthetic rating $r_{\mathrm{aes}}\in[0,1]$. 
To avoid rewarding visually appealing but incorrectly implemented slides, we gate aesthetics by implementation completeness:
\begin{equation}
    r_{\mathrm{sty}} \;=\; r_{\mathrm{imp}} \times r_{\mathrm{aes}} .
\end{equation}
Accordingly, the designer reward is defined as
\begin{equation}
    R_D \;=\; \alpha \, r_{\mathrm{cmp}} \;+\; \beta \, r_{\mathrm{cpl}} \;+\; \gamma \, r_{\mathrm{sty}},
\end{equation}
where $\alpha,\beta,\gamma \ge 0$ are scalar weights.

This alternating optimization scheme ensures coordinated learning between the designer and coder models. 

\begin{table*}[htbp]
\centering
\small
\setlength{\tabcolsep}{1.5mm}
\begin{tabular}{lcccccccc}
\toprule
\multirow{2}{*}{\textbf{Method}}
& \multicolumn{2}{c}{\textbf{Objective Metrics}}
& \multicolumn{2}{c}{\textbf{Content}} 
& \multicolumn{3}{c}{\textbf{Visual}} 
\\
\cmidrule(lr){2-3} \cmidrule(lr){4-5} \cmidrule(lr){6-8}
& \makecell{Success Rate (\%)}
& \makecell{Balance ($\uparrow$)}
& \makecell{Clarity}
& \makecell{Coherence}
& \makecell{Layout}
& \makecell{Hierarchy}
& \makecell{Color} 
& \makecell{Avg.} 
\\
\midrule
Direct${}_{Claude~Haiku}$      & \underline{99}  & 0.53 & \textbf{3.99} & 3.64 & 2.64 & 3.87 & 3.14 & 3.46 \\
Direct${}_{Claude~Sonnet}$     & 97  & 0.67 & \underline{3.79} & 3.47 & 2.67 & 3.64 & 3.13 & 3.34 \\
Auto-Slides${}_{Claude~Haiku}$ & 87  & \textbf{0.81} & 3.53 & 3.64 & 2.49 & 3.50 & 3.08 & 3.25 \\
EvoPresent${}_{Claude~Haiku}$  & \textbf{100} & 0.77 & 3.53 & \textbf{3.84} & \underline{3.46} & \underline{3.95} & \underline{3.69} & \underline{3.69} \\
\midrule
\textbf{DeepSlides${}_{Claude~Haiku}$} & 95  & \underline{0.78} & \underline{3.79} & \underline{3.81} & \textbf{3.55} & \textbf{4.04} & \textbf{3.71} & \textbf{3.78} \\
\bottomrule
\end{tabular}
\caption{We compare baseline systems and DeepSlides variants using objective metrics and VLM-as-judge scores for content quality  and visual design. Best and second-best results are highlighted in \textbf{bold} and \underline{underline}, respectively.}
\label{tab:main-results}
\vspace{-4mm}
\end{table*}



\section{Experiment}

Our experiments aim to address the following three questions:
(1) Does DeepSlides effectively accomplish slide design tasks while maintaining high-quality content?
(2) Does decoupling the design and implementation processes enhance the aesthetic quality of the slides?
(3) How is the performance of DeepSlides perceived by human evaluators?

\subsection{Baselines}
\label{sec:exp-setup}

To evaluate DeepSlides, we conducted tests using claude-haiku-4.5~\cite{claude-haiku-4.5} as foundational model. We compared our results against slides directly generated by claude-haiku-4.5 and claude-sonnet-4.5~\cite{claude-sonnet-4.5}. Additionally, we benchmarked our approach against two prominent slide generation workflows: EvoPresent~\cite{liu2025presentingisanart} and Auto-Slides~\cite{yang2025autoslides}, with claude-haiku-4.5 as the base model. During the user requirement simulation phase, we utilized gpt-4o-mini~\cite{openai-gpt4o-mini} as the foundational model, and the generated research reports served as inputs for the other two workflows.

\subsection{VLM as a Judge}

We employ two categories of metrics to comprehensively assess slide design quality: objective metrics (Success Rate, Balance) and Vision-Language Model (VLM)-derived metrics. \textit{Success Rate} measures the fraction of generations that can be successfully rendered without errors, while \textit{Balance} quantifies the spatial evenness of the layout distribution. The VLM-derived metrics include content dimensions (Clarity, Coherence) and visual dimensions (Layout, Hierarchy, Color Scheme), assessed using GPT-5.2~\cite{openai_gpt52_system_card_2025}. Detailed definitions are provided in Appendix~\ref{sec:metrics}.

As summarized in Table~\ref{tab:main-results}, {DeepSlides} exhibits superior overall performance relative to baseline approaches, obtaining the highest average VLM-based evaluation score (3.78). While the Success Rate of DeepSlides (95\%) slightly trails EvoPresent (100\%), DeepSlides notably surpasses EvoPresent and other baselines (Direct, Auto-Slides) in crucial content metrics such as Layout (3.55 vs. 3.46), Hierarchy (4.04 vs. 3.95), and Color Scheme (3.71 vs. 3.69). 
Auto-Slides attains a higher Balance score than DeepSlides primarily because it generates slides via \LaTeX, where the layout is largely fixed and thus exhibits less spatial variation across outputs, leading to more stable balance measurements.
These findings confirm that DeepSlides effectively accomplishes slide design tasks without compromising content quality.

\subsection{Human Evaluation}
To assess DeepSlides' alignment with human preferences, we recruited 20 volunteers to rank slides generated on identical topics across frameworks, with the slide order randomly shuffled to prevent presentation bias.
We randomly sample 40 topics for human evaluation and use Claude Haiku 4.5 and Claude Sonnet 4.5 as base models (20 topics per model). The evaluators are split into two groups of ten, with one group completing the ranking task and the other completing the scoring task.

We report the frequency of first-place selections and the pairwise win rates of our framework against each competing method. Additionally, evaluators rated DeepSlides' slides across four dimensions. Detailed evaluation criteria and questionnaires are presented in Appendix~\ref{sec:human_eval}.

\begin{figure}[ht]
  \centering  \includegraphics[width=1.0\linewidth]{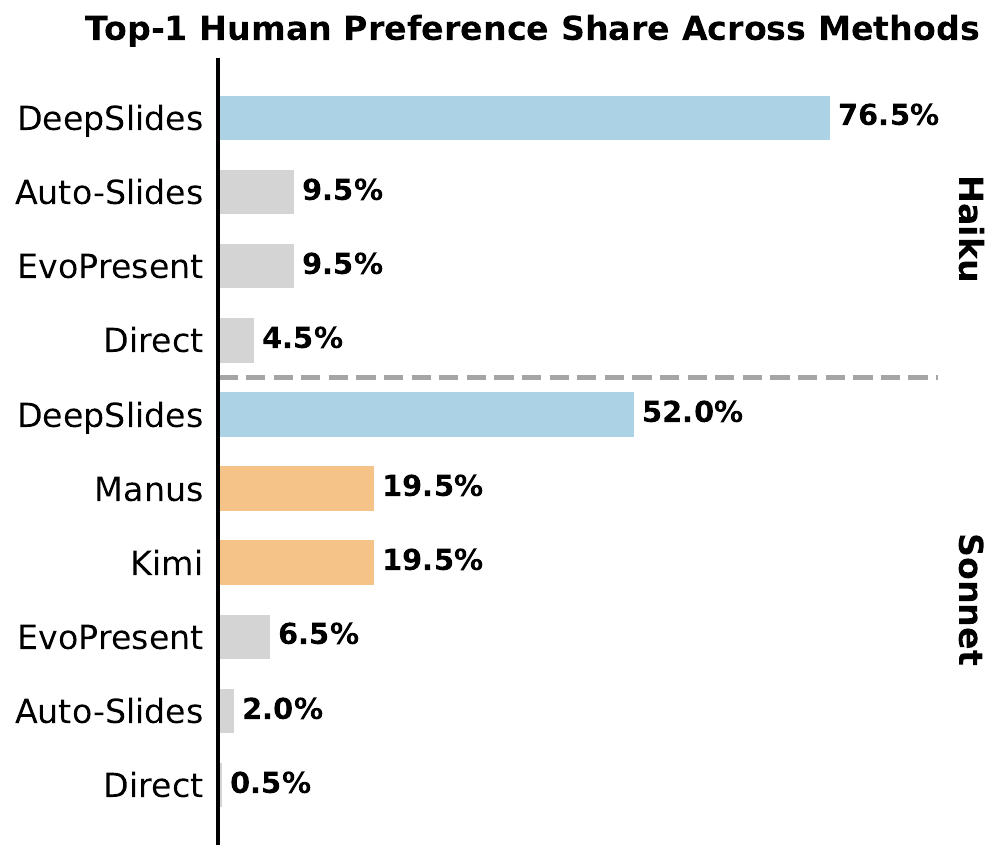}
  \caption{We report the fraction of topics for which each framework is ranked \#1 by human evaluators.}

  \label{fig:rank_1}
  \vspace{-4mm}
\end{figure}

\paragraph{DeepSlides is most frequently ranked as the top method in human evaluation.} As shown in Figure~\ref{fig:rank_1}, it achieves the highest Top-1 preference rate under both Claude Haiku 4.5 and Claude Sonnet 4.5 settings, with its Top-1 cumulative preference consistently exceeding all competing approaches. This indicates that human raters most often select DeepSlides as their first-choice system, reflecting stronger overall preference relative to the baselines.

\begin{figure}[ht]
  \centering  \includegraphics[width=1.0\linewidth]{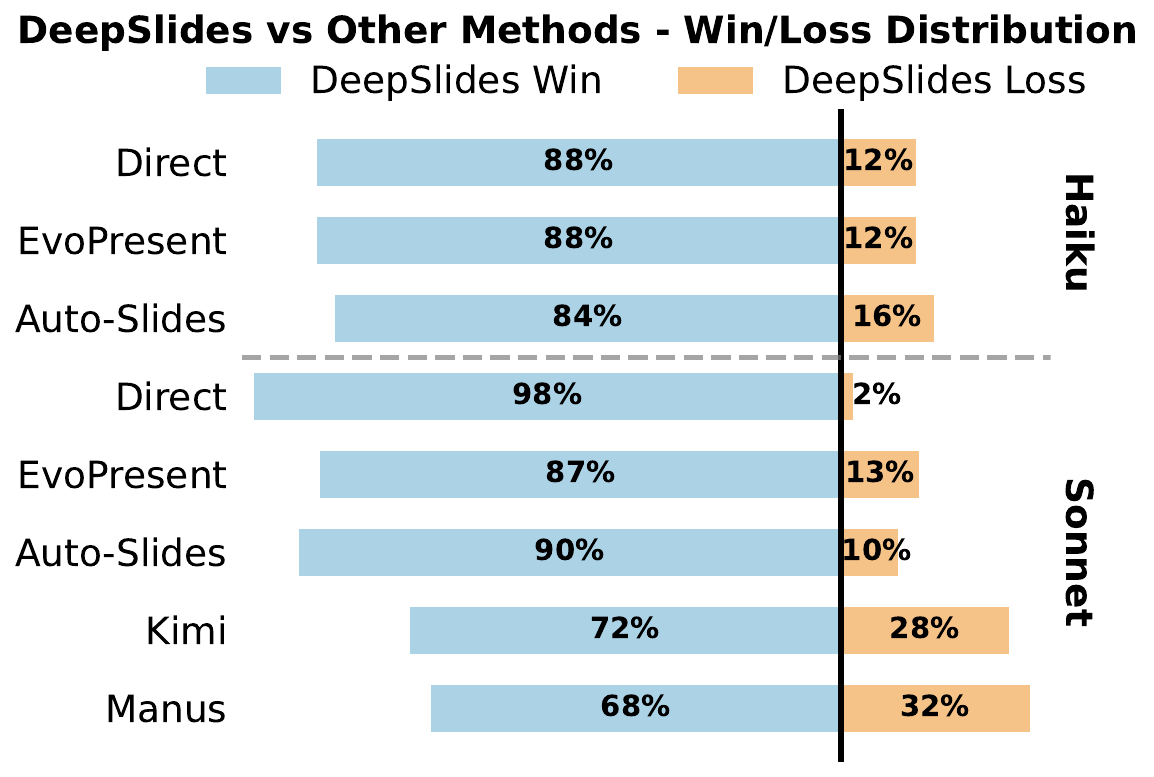}
  \caption{Pairwise win rate.}
  \label{fig: onebyone}
  \vspace{-4mm}
\end{figure}

\paragraph{DeepSlides is consistently preferred by human evaluators over all baselines. } As shown in Figure~\ref{fig: onebyone}, DeepSlides achieves a dominant win rate against all baselines under both Claude Haiku 4.5 and Claude Sonnet 4.5 settings, with only a small fraction of ties or losses. Notably, it remains strongly preferred even when compared to commercial proprietary solutions (e.g., Kimi and Manus), indicating that the proposed approach delivers superior perceived quality beyond open-source and template-based alternatives.

\begin{figure}[ht]
  \centering  \includegraphics[width=0.9\linewidth]{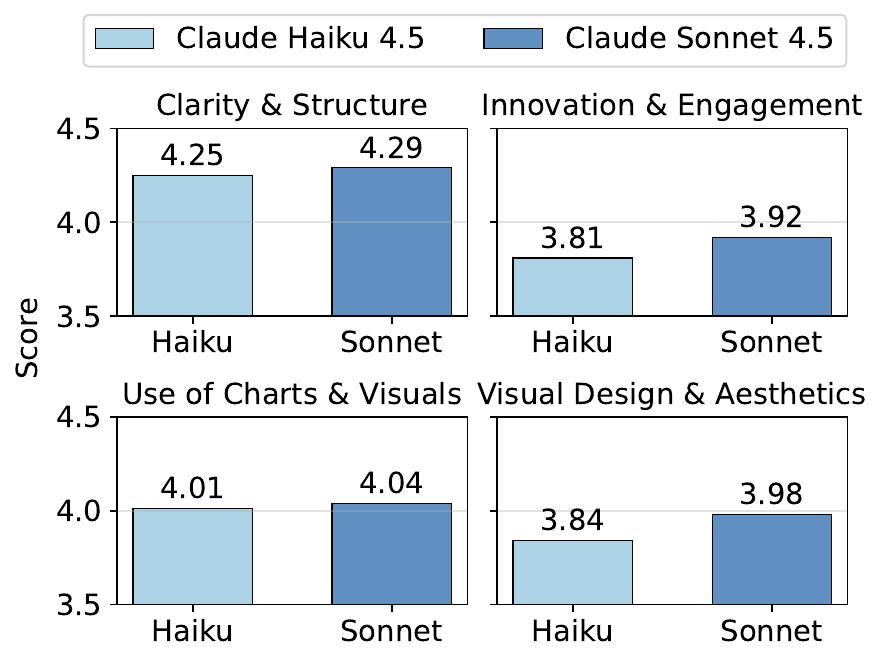}
  \caption{Human evaluation scores.}
  \label{fig: human_rating}
  \vspace{-4mm}
\end{figure}

\paragraph{DeepSlides Achieves High Scores in Human Evaluation.} As shown in Figure \ref{fig: human_rating}, human evaluators provided detailed ratings across various dimensions. DeepSlides received strong ratings for \textit{Clarity \& Structure}, achieving scores of 4.25 for Haiku and 4.29 for Sonnet, indicating excellent content organization. Ratings for \textit{Innovation \& Engagement} and \textit{Use of Charts \& Visuals} are also notable, demonstrating DeepSlides' capability to produce engaging and visually informative slides. Furthermore, the high scores in \textit{Visual Design \& Aesthetics} underline the effectiveness of the proposed decoupling strategy in enhancing visual appeal.





\begin{table}[htbp]
\small
    \centering
    \begin{tabular}{lccc}
        \toprule
        Model & Success Rate & Balance & Visual \\
        \midrule
        Qwen3-0.6B & 0.0 & -  & - \\
        SlideQwens${}_{sft}$  & 81.4 & \textbf{0.77} & 3.23 \\
        SlideQwens${}_{full}$  & \textbf{86.0}  & \textbf{0.77} & \textbf{3.38} \\
        \bottomrule
    \end{tabular}
\caption{Model Training Performance Comparison.}
\label{tab:ablation}
\vspace{-4mm}
\end{table}

\subsection{Training Results}

To validate our training strategy, we evaluate the base, SFT, and SFT+RL models on single-slide generation, where each model must jointly produce a layout and executable code given page content and randomly sampled design guidelines. As shown in Table~\ref{tab:ablation}, the base model (\textit{Qwen3-0.6B}) fails to complete the task, whereas SFT enables \textit{SlideQwens${}{sft}$} to reach an $81.4\%$ success rate (Balance $0.77$, Visual $3.23$). Adding MARL further improves reliability and visual quality: \textit{SlideQwens${}{full}$} achieves $86.0\%$ success rate and higher Visual score ($3.38$) with comparable Balance ($0.77$), indicating that SFT is necessary for executability while MARL yields additional quality gains.
\section{Conclusion}

In this paper, we propose \textbf{DeepSlides}, a hierarchical workflow for automated slide generation that explicitly decouples slide design from implementation, addressing the limitations of prior work that either rely on fixed templates or bypass the essential design stage. We curate \textbf{SlideDesign}, a dataset tailored for slide‑page design and implementation, and introduce a multi‑agent reinforcement learning training paradigm to train the \textbf{SlideQwens} model family. Extensive empirical evaluations, including objective metrics, vision‑language model judgments, and human preference studies, demonstrate that DeepSlides achieves superior completeness, compliance, and aesthetic quality compared with competitive baselines. Our ablation studies further confirm that staged supervised fine‑tuning followed by reinforcement learning is critical for balancing content quality and visual attractiveness in structured multimodal slide tasks. Collectively, these findings validate that decoupling design and implementation within a structured workflow substantially improves automatic slide generation, producing outputs that better align with human evaluators in both content and visual design.

\section*{Acknowledgement}
This research was supported by the National Natural Science Foundation of China (No. U22B2036), the National Natural Science Foundation of China (No. 62506186), the Technological Innovation Team of Shaanxi Province (No. 2025RS-CXTD009), and the International Cooperation Project of Shaanxi Province (No. 2025GH-YBXM-017).
\section*{Limitations}
Despite improved controllability via decoupling design from implementation and feedback-driven specialization, our method has several limitations. The multi-stage pipeline increases inference latency and operational complexity compared to single-pass generation. While we aim for template-free design, achieving consistent brand-level styling (e.g., strict typography and color constraints) may still require user-provided style references or stronger conditioning, which we do not systematically study. Moreover, our automatic evaluators provide only a proxy for human judgments of aesthetics and communicative effectiveness, leading to failure modes such as over-decoration, suboptimal whitespace, and occasional divergences between design intent and rendered output. Finally, experiments are conducted primarily in academic slide settings with a fixed set of rendering backends; broader generalization to other domains and authoring environments remains an open question.

From a deployment perspective, the system entails additional risks. It may inadvertently violate intellectual property or licensing constraints in text and visual assets without explicit provenance tracking and license-aware filtering. Because the pipeline emits executable slide code and relies on external rendering backends, it also introduces security considerations (e.g., prompt injection and unsafe HTML/SVG/code paths), necessitating sandboxing and rigorous sanitization. Additionally, learned design priors may encode cultural bias or undermine accessibility, and the ability to generate high-quality persuasive decks increases misuse potential, motivating policy enforcement and auditing in practical deployments.

\section*{Ethics Statement}
Human experts participated voluntarily in providing evaluation ratings, and no personally identifiable or sensitive data were collected; all procedures complied with institutional ethical guidelines.

\bibliography{custom}

\appendix
\clearpage



\section{Prompts}
\label{sec:prompts}

\begin{tcolorbox}[title={Prompt for Designing Slide}]
You are a seasoned slide designer responsible for designing slide layouts.
Generate a slide layout description in JSON format based on the following details:

Title: <slide title>

Detailed points: <Main points>

<Slides features>

<Style Instructions>

<Layout Instructions>

Here are some images you may optionally use in the PPT:
<Image list>

<Suggestions for improving design>

OUTPUT: <Design format>

\end{tcolorbox}

\begin{tcolorbox}[title={Prompt for Coding}, breakable]
Generate Python code that creates slides using the python-pptx library based on the following detailed slide description:

Title: <slide title>

Detailed points: <Main points>

Design Details: <Agent's Design for the Slide>

<Image list>

<Tool list>

Code requirements:

1. Import the necessary libraries.

2. Create the slides and ensure the widescreen standard aspect ratio: 16:9 (13.33 inches × 7.5 inches).

3. According to the detailed description, add the title, bullet points, and images at specified positions; set fonts and styles; explicitly set the size of each element to prevent overlap/occlusion; ensure text wraps automatically. 

4. Only the provided image URLs can be used. Do not reserve any positions for any images that are not provided, and do not use text descriptions to fill the gaps. Or you can also manually create some flowcharts using various graphics, but don't just leave an empty space or just provide a textual description.

5. All the text should be placed on the top layer.

6. Save the file as: <Save target>

Previous code and errors (if any):
<previous code>
<error message>

Please provide complete, executable Python code based on this information. 

Note: output Python code only, do not output any other text.
Code will be save in utf-8 encoding.
\end{tcolorbox}

\section{Dataset}
\label{sec:dataset}

We allocate 10 topics per secondary field (420 topics total). 
Our released dataset only contains (i) source webpage URLs and (ii) concise webpage summaries, and does not redistribute raw webpage content.
To mitigate potential privacy and safety risks, we apply automatic screening to remove or mask personally identifying information (e.g., email addresses, phone numbers, and person names when present) from summaries, and filter samples using keyword- and model-based checks for potentially offensive content.
We additionally perform manual spot-checking on a subset of the collected entries and discard flagged items.
We randomly select 50 topics for supervised fine-tuning (SFT) cold-start data, 200 topics for reinforcement learning (RL) training data and 100 topics for the main experiments. We randomly select 500 pages of content for the ablation study.
For generating SFT data, the pipeline employs layout design and Python code generation steps utilizing the \texttt{gpt-5} model (designer) and \textit{claude-haiku-4.5} model (coder). Outputs undergo a rigorous check against quality thresholds, resulting in a final dataset of 1,286 training pairs.

\section{Experiment Details}
\label{sec:exp_details}

\subsection{Training Details}
\label{sec:training_details}

\paragraph{SFT Cold Start.}
We conduct supervised fine-tuning (SFT) of a Qwen3-0.6B model using full-parameter training . We use AdamW with a learning rate of $1\times10^{-4}$, cosine scheduling, weight decay of 0.01, and gradient clipping at 1.0. The per-device batch size is 1 with 8-step gradient accumulation (effective batch size 8). We train for 10 epochs with a warm-up ratio of 0.1, using bf16 and gradient checkpointing.

\paragraph{MARL.}
We train the MARL policy with a GRPO-based pipeline using a Qwen3-0.6B backbone. We set the train/validation batch size to 8 and cap both prompt and response lengths at 8{,}000 tokens. Rollouts use single-turn generation with $n=10$ samples per prompt. GRPO uses a learning rate of $1\times10^{-6}$ with mini-batch size 2 and micro-batch size 4 per GPU. We enable a switch-agent mechanism every 10 steps between two specialized agents. Training runs for 10 epochs (500 steps), with evaluation every 100 steps and checkpointing every 50 steps.

All training was conducted on NVIDIA H200 GPUs; supervised fine-tuning used 2 GPU-hours, and RL training used 400 GPU-hours (402 GPU-hours in total).

\subsection{Metrics}
\label{sec:metrics}

\paragraph{Balance.}
We define a visual-weight map $w(x,y)$ on a grayscale image $I$ as a mixture of background-relative deviation and local contrast:
\[
\begin{aligned}
w(x,y) &= (1-\lambda)\,|I(x,y)-b| \\
       &\quad + \lambda\,\bigl|I(x,y)-(\mathcal{G}_\sigma * I)(x,y)\bigr|.
\end{aligned}
\]

where $b=\mathrm{median}(I)$ and $\mathcal{G}_\sigma$ is a Gaussian blur. Let the total weight be $M=\sum_{x,y}w(x,y)$.

The weight center of mass $(c_x,c_y)$ yields a centeredness score
\[
s_{\mathrm{com}} = 1 - \frac{1}{\sqrt{2}}\sqrt{\Bigl(\frac{|c_x-W/2|}{W/2}\Bigr)^2 + \Bigl(\frac{|c_y-H/2|}{H/2}\Bigr)^2}.
\]

Define left/right and top/bottom weight sums $M_L,M_R,M_T,M_B$. Then
\[
s_{\mathrm{lr}} = 1 - \frac{|M_L-M_R|}{M},\qquad
s_{\mathrm{tb}} = 1 - \frac{|M_T-M_B|}{M}.
\]

The final balance score is the average of these three components,
\[
\mathrm{Balance} = \mathrm{clip}_{[0,1]}\!\left(\frac{s_{\mathrm{com}}+s_{\mathrm{lr}}+s_{\mathrm{tb}}}{3}\right).
\]

\paragraph{Clarity.} Assess how easily the information can be understood at a glance, focusing solely on the content's clarity and logical structure. This includes the clarity of language, how well the information is explained, and whether the key points are immediately apparent. A high score means the content is direct, concise, and easy to understand without unnecessary complexity.

\paragraph{Coherence.} Evaluate how logically the content is organized. A clear narrative flow is critical, with well-structured sections and seamless transitions between ideas. We encourage the use of both text and visuals to enhance clarity and support the narrative. A high score means the content is organized in a way that guides the viewer through the slide’s message and is complemented effectively by visual elements.


\paragraph{Layout.} Analyze the alignment, spacing, and overall balance of the slide’s elements. Creative and thoughtful typography and element placement enhance the visual appeal and engagement. Single-block content layouts are discouraged, as they tend to be overly monotonous and can fail to capture the audience's interest. Additionally, the slide should incorporate tasteful decorative elements that enhance the overall design without overwhelming the content. A high score reflects a well-structured, balanced slide with sufficient white space, and ideally some creative flair in the layout design.

\paragraph{Hierarchy.} This metric assesses whether there is a clear visual distinction between primary and secondary information. It evaluates if the audience can immediately identify titles, main points, and supporting details. This is typically achieved through variations in font size, weight (boldness), color, and placement.

\paragraph{Color Scheme.} Evaluate how rich and aesthetically pleasing the color palette is. This includes assessing the use of a variety of colors in a harmonious way, with good contrast that enhances readability and supports the slide's overall theme. The slide should also feature some decorative elements with colors used thoughtfully and reasonably, contributing to both visual appeal and a balanced design. A high score means the colors are vibrant, well-coordinated, and support the slide's aesthetic and thematic cohesion. The color of the slides should correspond to the content.

\subsection{Other Details}
For all LLM calls, we set \texttt{temperature}=0 and \texttt{max\_tokens}=8192; all other parameters follow the default settings of the corresponding API/package.

\section{Human Evaluation}
\label{sec:human_eval}
\begin{figure*}[ht] 
    \centering
    \small 
    \begin{tcolorbox}
        
        \begin{center}
            \bfseries\large Questionnaire for Slides Ranking
        \end{center}
        \vspace{6pt}

        \textbf{Task:} 
            The slide decks below were generated based on an identical topic. Please conduct a comparative assessment of their overall quality (e.g., content clarity, structure, coherence, and visual presentation) and provide a complete ranking from the best-performing deck to the weakest.
        \vspace{2pt}
        \hrule height 0.3pt \vspace{5pt}

        \textbf{GROUP:} 1
        
        \textbf{TOPIC:} Dispute Resolution Innovation

        \begin{center}
        \includegraphics[width=0.9\linewidth]{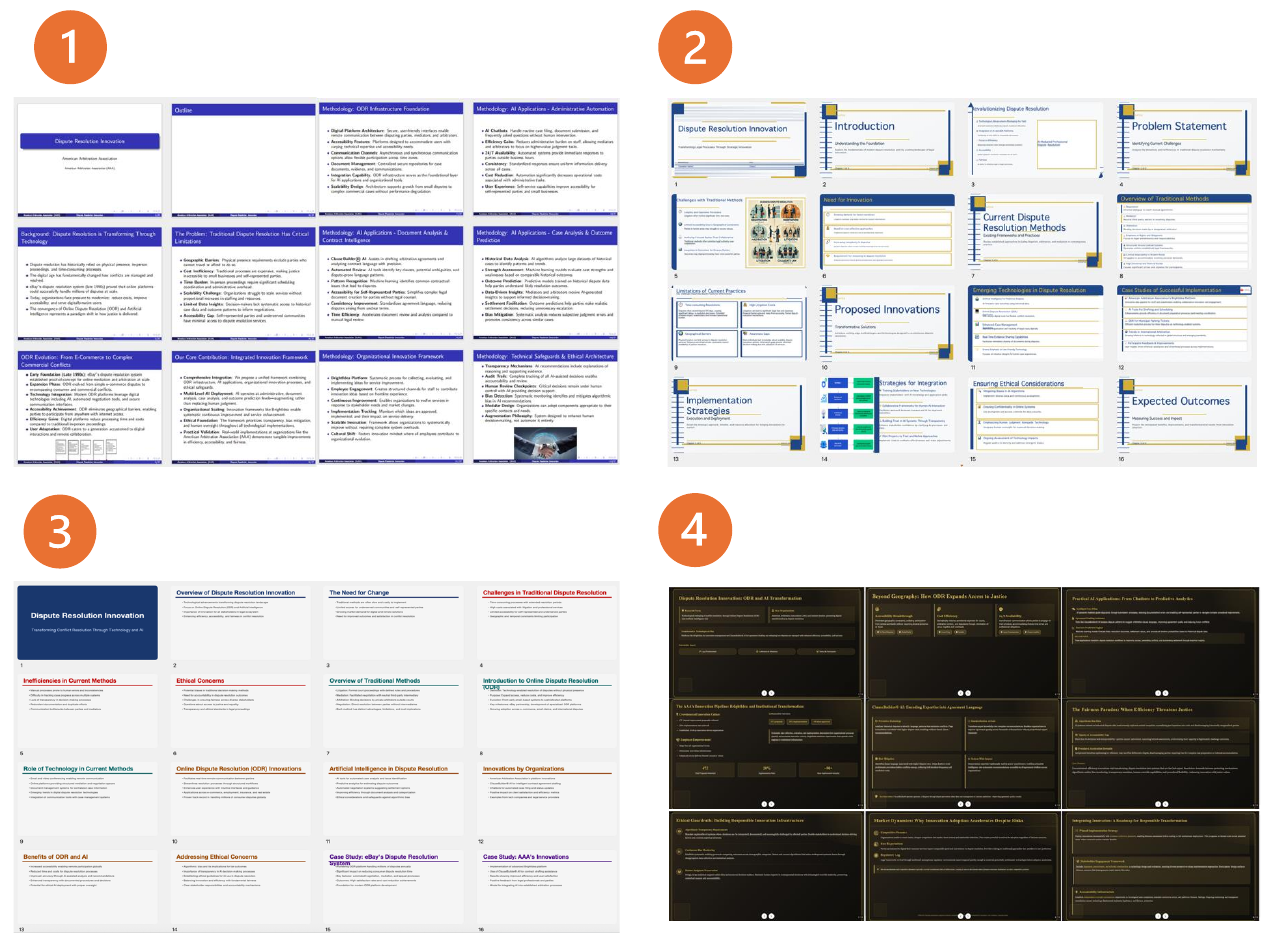}
        \end{center}

        \textbf{Please enter your ranking (from best to worst):}\ \underline{\hspace{6cm}}
        

    \end{tcolorbox}
    \caption{Detailed information on the human evaluation questionnaire.} 
    \label{instruction1}
\end{figure*}

\begin{figure*}[ht] 
    \centering
    \small 
    \begin{tcolorbox}
        
        \begin{center}
            \bfseries\large Questionnaire for Slides Evaluation
        \end{center}
        \vspace{6pt}

        
        \textbf{TOPIC:} Dispute Resolution Innovation

        \begin{center}
        \includegraphics[width=0.9\linewidth]{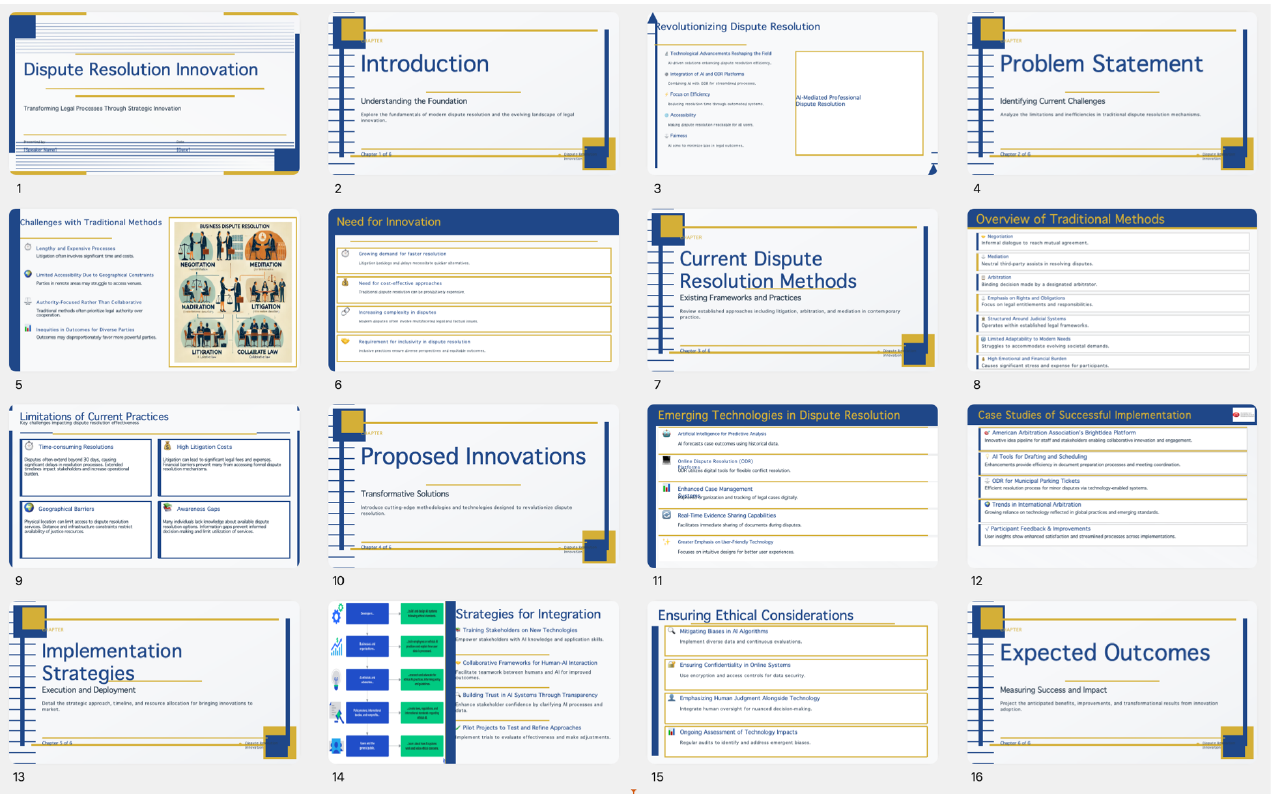}
        \end{center}

        
        \textbf{Task:} 
        \noindent Please rate the slide deck quality across the following four dimensions.
        \vspace{-6pt}
        \begin{enumerate}
            \item \textbf{Clarity \& Structure} \\
            \vspace{-12pt}
            \begin{itemize}
                \item[$\square$] \textbf{5}: The content is highly clear and logically rigorous, with a well-organized structure. 
                \item[$\square$] \textbf{4}: The content is clear and the structure is generally sound, though certain parts may appear slightly cluttered or somewhat redundant.
                \item[$\square$] \textbf{3}: The content is reasonably clear, but the structure is insufficiently explicit.
                \item[$\square$] \textbf{2}: The content lacks clarity; information is overly dispersed or difficult to comprehend.
                \item[$\square$] \textbf{1}: The content is disorganized with poor structure, making it hard to follow.
            \end{itemize}
        
            \item \textbf{Use of Charts \& Visuals} \\
            \vspace{-12pt}
            \begin{itemize}
                \item[$\square$] \textbf{5}: Charts and visual elements are used appropriately and effectively, substantially aiding explanation and comprehension.
                \item[$\square$] \textbf{4}: Charts and visuals are used well overall, but may occasionally be redundant or not sufficiently precise.
                \item[$\square$] \textbf{3}: Charts and visuals are limited, or in some places do not provide meaningful support for understanding.
                \item[$\square$] \textbf{2}: Charts and visuals are largely absent or potentially misleading.
                \item[$\square$] \textbf{1}: No charts or visual elements are provided, or the visuals are entirely irrelevant to the content.
            \end{itemize}
        
            \item \textbf{Innovation \& Engagement} \\
            \vspace{-12pt}
            \begin{itemize}
                \item[$\square$] \textbf{5}: The design and content are highly engaging and demonstrate strong creativity, effectively capturing audience attention.
                \item[$\square$] \textbf{4}: The presentation shows a degree of innovation and generally sustains audience interest.
                \item[$\square$] \textbf{3}: The presentation is relatively conventional, with only moderate innovation; audience engagement may not be sustained.
                \item[$\square$] \textbf{2}: The presentation lacks novelty and may lead to reduced audience interest.
                \item[$\square$] \textbf{1}: The presentation is entirely unoriginal; both content and design are monotonous and unengaging.
            \end{itemize}
        
            \item \textbf{Visual Design \& Aesthetics} \\
            \vspace{-12pt}
            \begin{itemize}
                \item[$\square$] \textbf{5}: The visual design is outstanding, with excellent color harmony, well-varied layout, and effective integration of text and images.
                \item[$\square$] \textbf{4}: The design is aesthetically pleasing overall, though some aspects could be further refined.
                \item[$\square$] \textbf{3}: The design quality is average; color choices and/or layout require improvement, and certain elements feel inconsistent.
                \item[$\square$] \textbf{2}: The design is cluttered, with incoherent color schemes and a confusing mix of text and images.
                \item[$\square$] \textbf{1}: The design is poor, with weak visual quality; color and layout are confusing and impede readability.
            \end{itemize}
        \end{enumerate}


        

    \end{tcolorbox}
    \caption{Detailed information on the human evaluation questionnaire.} 
    \label{instruction2}
\end{figure*}

We recruit 20 graduate students from diverse disciplines as human evaluators; participation is unpaid.
All human evaluators participated voluntarily with informed consent, and were informed that their annotations would be used solely for research purposes. Our human expert evaluators were aged 20--50, all had at least a university-level education, demonstrated strong English proficiency, and were based in Asia.
We store and report the collected ratings in anonymized and aggregated form. 
Our questionnaire is shown in Figure~\ref{instruction1} and Figure~\ref{instruction2}.

\section{License and Availability}
All released artifacts (including code, evaluation scripts, and accompanying resources) are distributed under the MIT License. We use all third-party artifacts (datasets, models, and toolchains) in accordance with their stated intended use and access conditions. Our released artifacts are intended for research and evaluation only; we do not redistribute any third-party data beyond what their original terms permit, and any derived resources remain subject to the same restrictions.

\section{Case Study}
\label{sec:examples}
\begin{figure*}[htbp]
  \centering
  \includegraphics[width=1.0\linewidth]{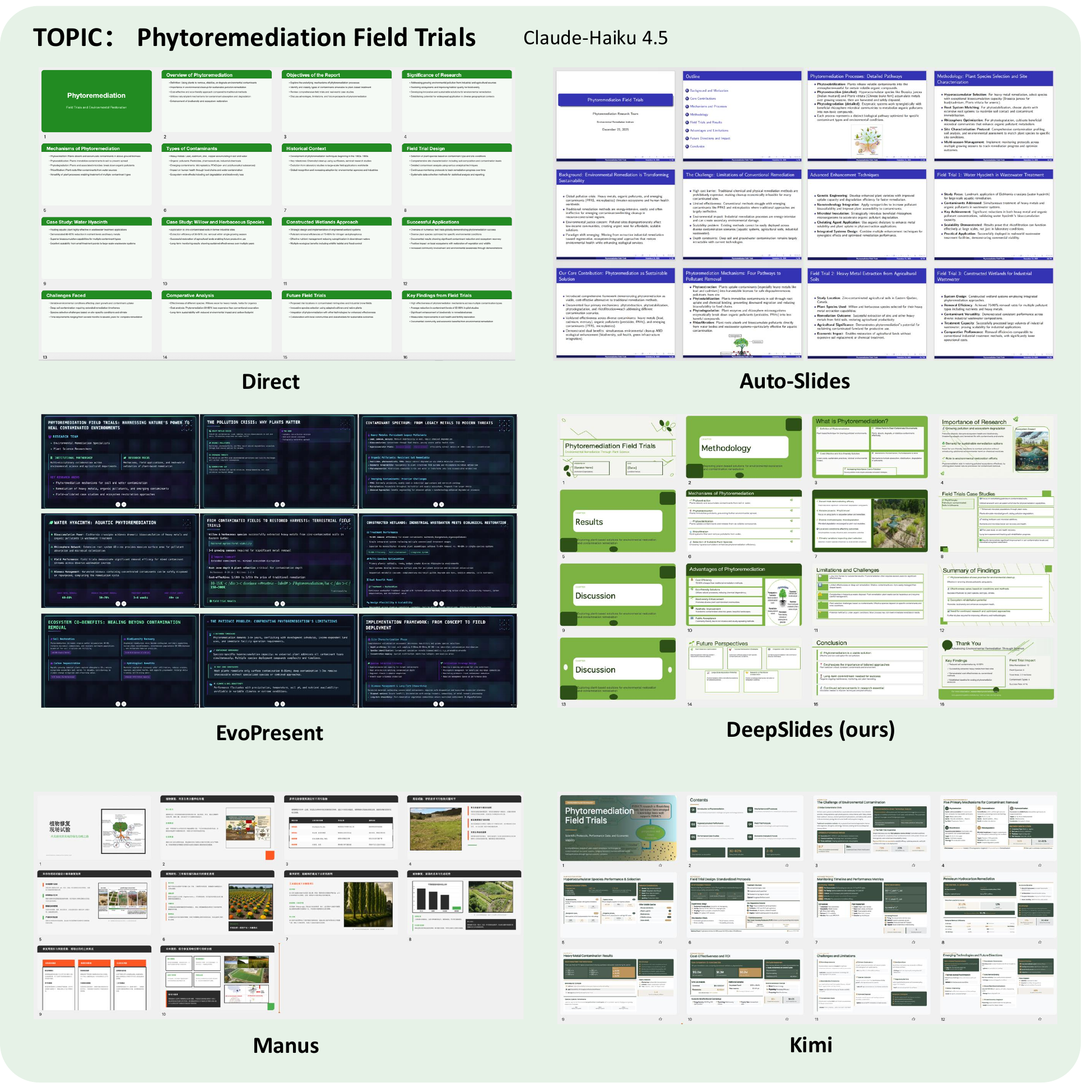}
  \caption{Case Study.}
  \label{fig:example1}
\end{figure*}

\begin{figure*}[htbp]
  \centering
  \includegraphics[width=1.0\linewidth]{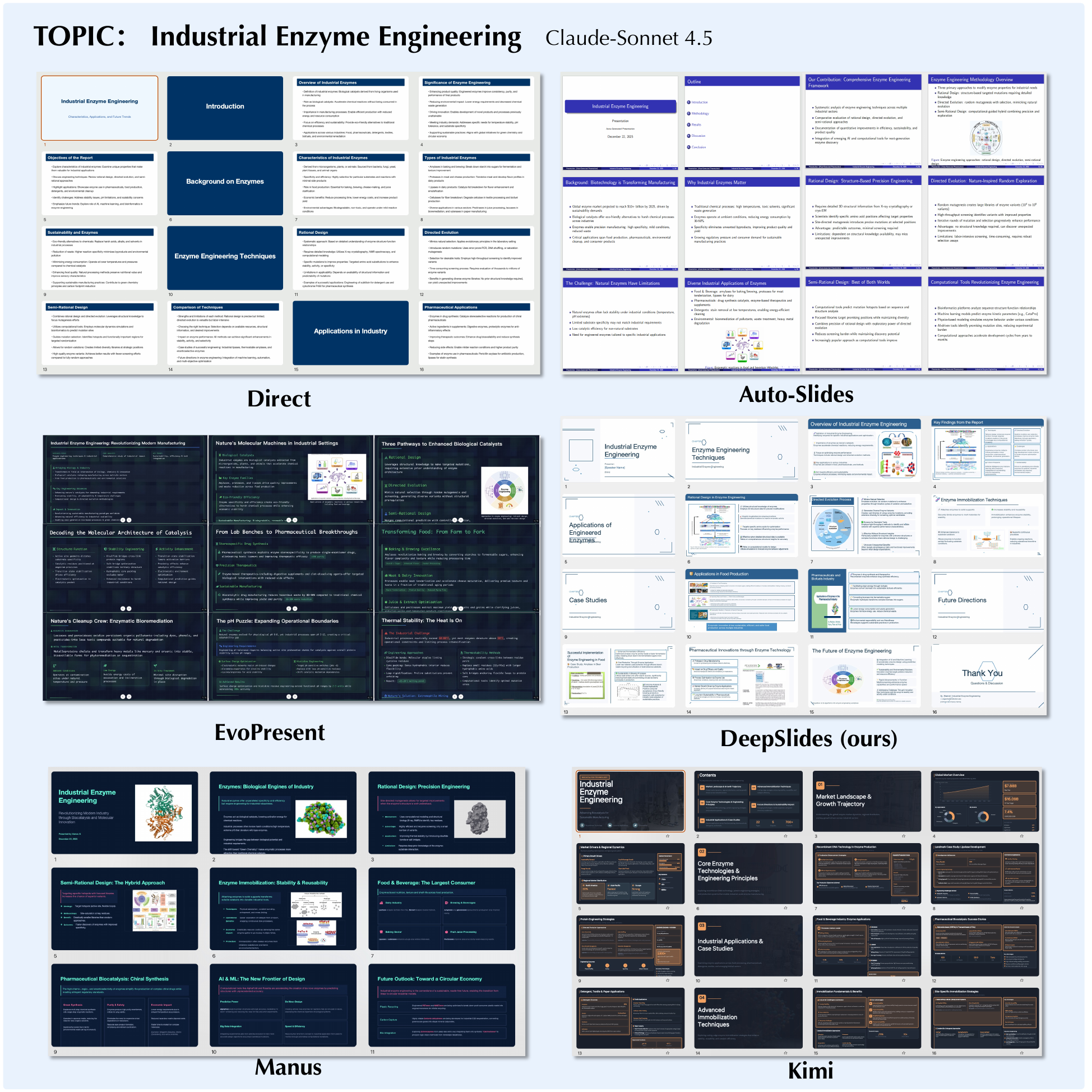}
  \caption{Case Study.}
  \label{fig:example5}
\end{figure*}

\begin{figure*}[htbp]
  \centering
  \includegraphics[width=\linewidth]{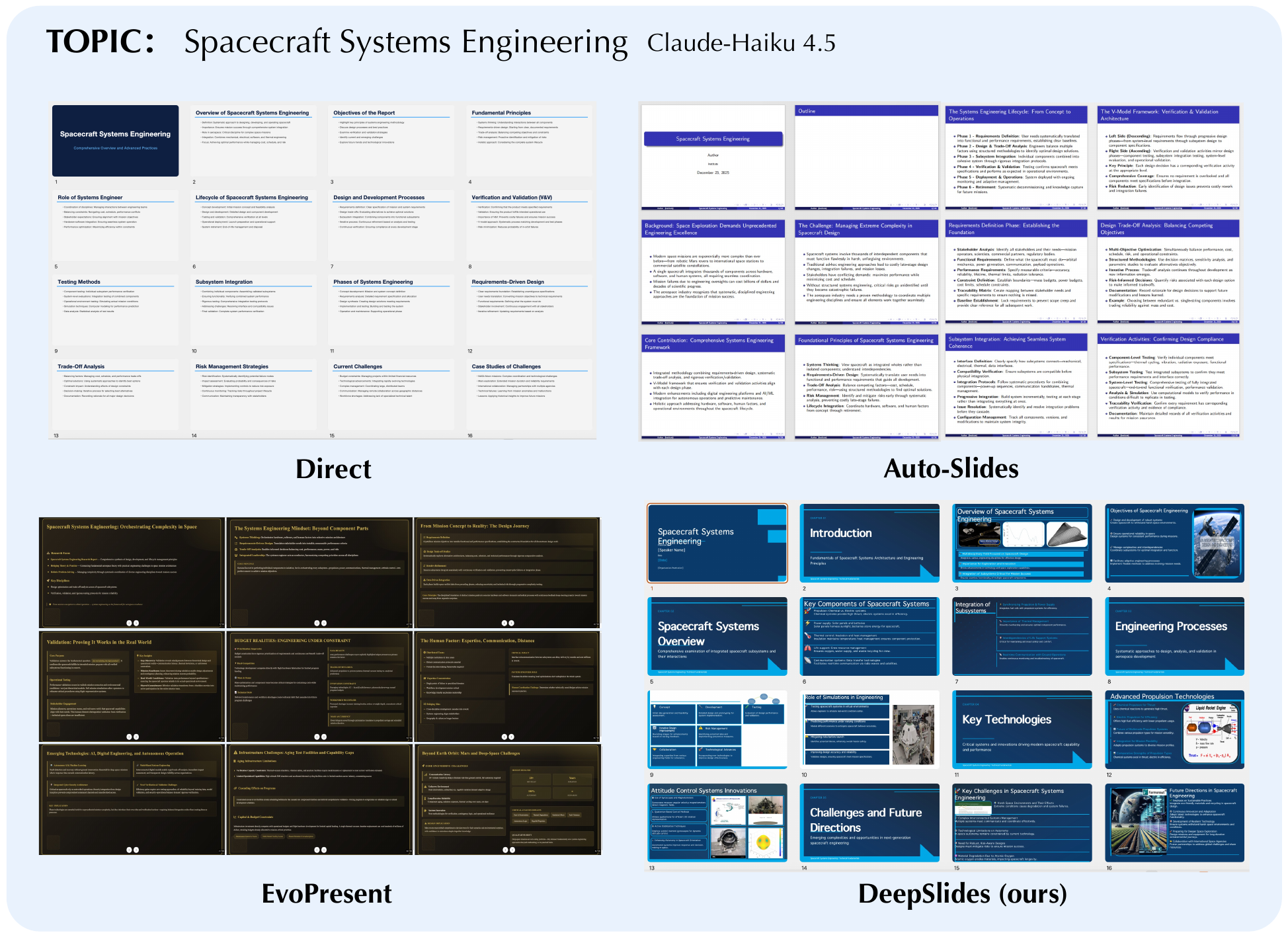}
  \caption{Case Study.}
  \label{fig:example4}
\end{figure*}

\begin{figure*}[htbp]
  \centering
  \includegraphics[width=1.0\linewidth]{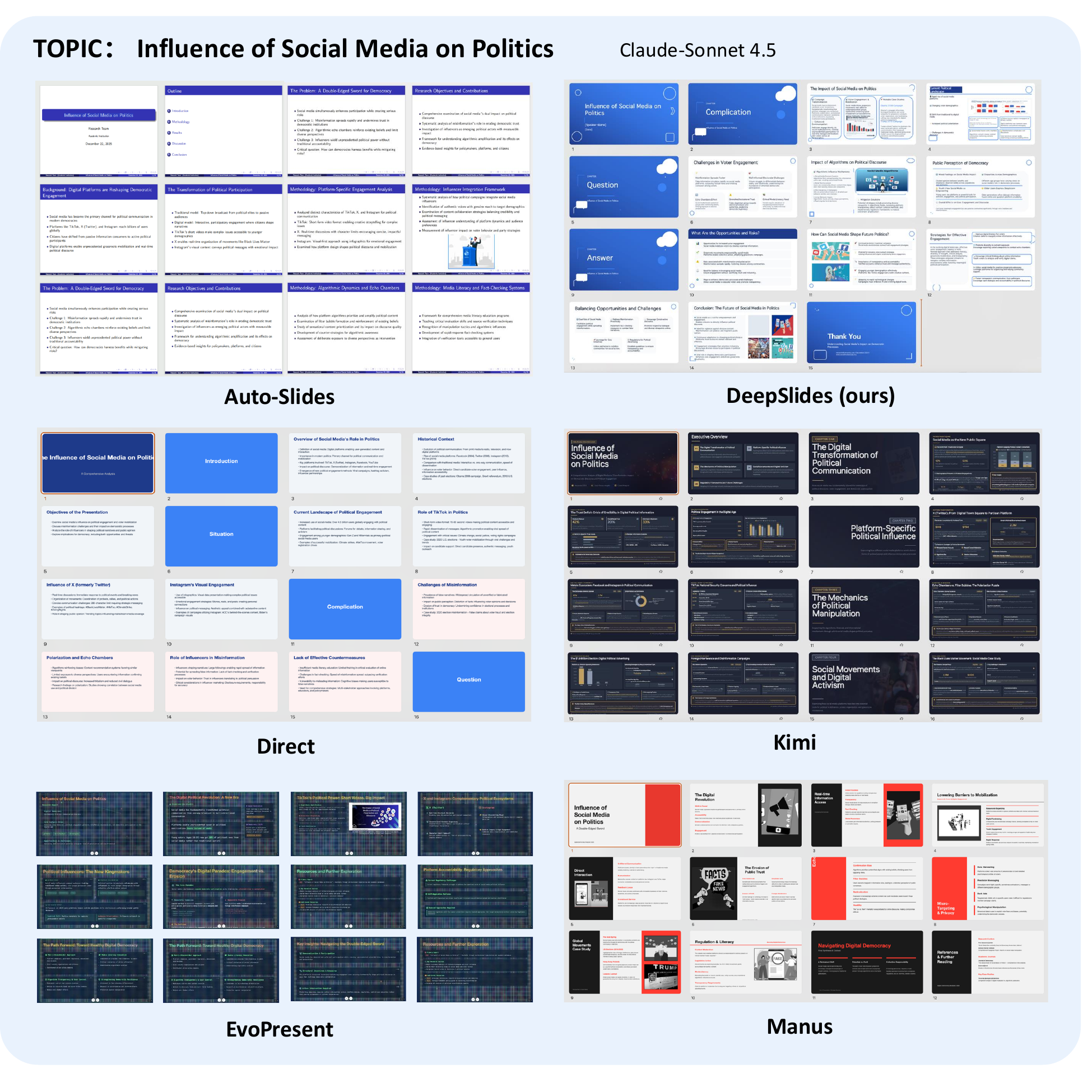}
  \caption{Case Study.}
  \label{fig:example2}
\end{figure*}

\begin{figure*}[htbp]
  \centering
  \includegraphics[width=1.0\linewidth]{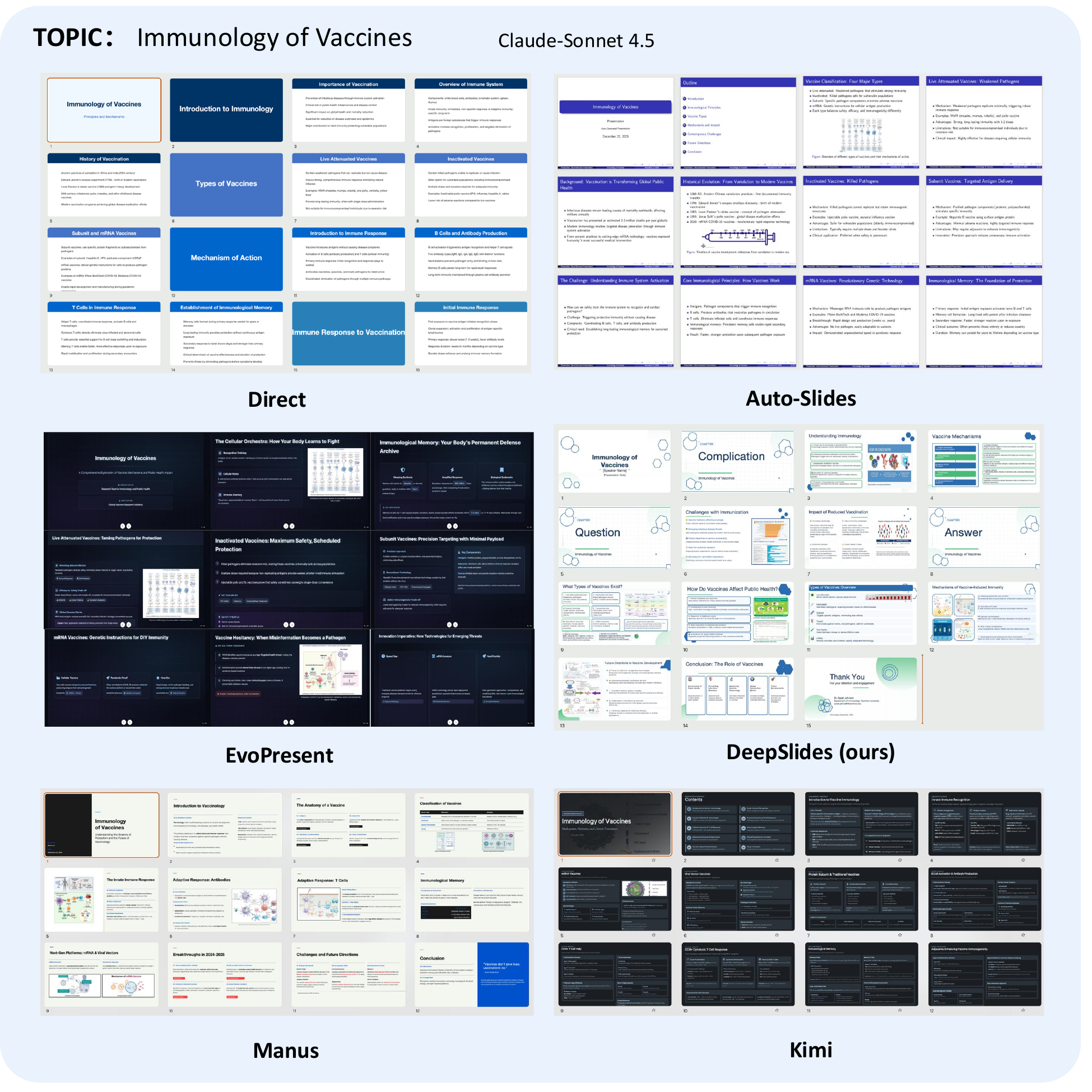}
  \caption{Case Study.}
  \label{fig:example3}
\end{figure*}

As shown in Figures~\ref{fig:example1}-\ref{fig:example3}, we present several comparative examples.

\clearpage

\end{document}